\documentstyle[aps,tighten,epsfig]{revtex}
\begin{document}
\draft
\title{
Evaluation of QCD sum rules for light vector mesons
at finite density and temperature}
\author{
{\sc S. Zschocke$^a$, 
O.P. Pavlenko$^{a,b}$, 
B. K\"ampfer$^a$}
}
\address{
$^a$ Forschungszentrum Rossendorf, PF 510119, 01314 Dresden, Germany \\
$^b$ Institute for Theoretical Physics, 03143 Kiev - 143, Ukraine 
}

\maketitle

\begin{abstract}
QCD sum rules are evaluated at finite nucleon densities and temperatures 
to determine the change of mass parameters 
for the lightest vector mesons $\rho$, $\omega$ and $\phi$ in a  
strongly interacting medium. For conditions relevant for the starting
experiments at HADES we find that the in-medium mass shifts of the
$\rho$ and $\omega$ mesons are governed, within the Borel QCD sum rule
approach, by the density and temperature dependence of the four-quark
condensate. In particular, the variation of the strength 
of the density dependence of the
four-quark condensate  reflects directly the decreasing
mass of the $\rho$ meson and can lead to a change of the sign 
of the $\omega$ meson mass shift as a function of the density.
In contrast, the in-medium mass of the $\phi$ meson
is directly related to the chiral strange quark condensate which seems
correspondingly accessible. \\[6mm]
\end{abstract}

\section{Introduction}

The in-medium modifications of the light vector mesons 
($\rho$, $\omega$, $\phi$)
receive growing attention both from theoretical and experimental sides.
On the theoretical side there are various indications concerning an important
sensitivity of vector mesons to partial restoration of chiral symmetry in a hot
and dense nuclear medium. In particular, at finite temperature 
the vector and axial-vector correlation functions, 
which are related to the meson spectral densities, become
mixed in accordance with in-medium Weinberg sum rules 
\cite{Weinberg}.
At low temperature this mixing can be expressed
directly via vacuum correlators in a model independent way
\cite{Elitsky_Joffe}.
Additionally, as shown within lattice QCD \cite{Karsch} and various
effective model calculations \cite{Brown_Rho}, 
the chiral quark condensate as order parameter
decreases with increasing temperature and baryon density. 
Within the QCD sum rule approach, considerable in-medium modifications 
of vector meson masses even at normal nuclear
matter density have been predicted \cite{Hatsuda_Lee,Hatsuda_Lee_Shiomi}.

On the experimental side there is the idea to probe the 
in-medium modifications of vector
mesons, in particular mass shifts, by measuring dileptons ($e^+ e^-$) 
from meson decays.
This is the primary motivation of the presently starting experiments 
with the HADES detector
\cite{HADES} at the heavy-ion synchrotron SIS at GSI Darmstadt. 
At higher beam energies the heavy-ion experiments of the CERES collaboration \cite{CERES}
evidenced already hints to a noticeable modification of the dilepton spectrum
which can be reproduced under the assumption of a strong melting of the $\rho$ meson
in a dense, strongly interacting medium at temperatures close to the chiral
transition \cite{Rapp_Wambach,Gallmeister}.

While numerous evaluations of QCD sum rules have been performed during the last decade,
the majority of them deal either with cold nucleon matter or a hot pion medium.
At the same time, to extract the wanted information on the behavior of in-medium mesons
via measurements in heavy-ion collisions at SIS energies one has to study the case
of finite baryon density \underline{and} temperature. In this paper we present
evaluations of the QCD sum rules for the light in-medium vector mesons
in a 
density - temperature region which is relevant for experiments like the ones at HADES.
We study systematically here the relative numerical contributions of the four-quark
condensate to the QCD Borel sum rule to find out the general trends of the vector
meson mass dependence on density and temperature with respect to variations of the
poorly known four-quark condensate.

Our paper is organized as follows. In section 2 we recapitulate the general structure
of the QCD Borel sum rule. The operator product expansion and evaluation
of the needed condensates are summarized in section 3.
In section 4 we discuss the form of the hadronic spectral density.
The numerical evaluation of the sum rule is presented in section 5.
The summary can be found in section 6. 

\section{QCD sum rules at finite density and temperature} 

Following the approach developed in \cite{BS,Hatsuda_Koike_Lee} we employ for the QCD sum
rules (QSR) at finite nucleon chemical potential $\mu_N$ and temperature $T$ the
retarded current-current correlation function in a medium
\begin{equation}
\Pi^R_{\mu \nu} (q; \mu_N, T) = i \int d^4 x \;   
{\rm e}^{i \, q  x} \;
\langle {\cal R} J_{\mu}(x) J_{\nu}(0) \rangle_{\mu_N, T},
\label{eq_1}
\end{equation}
where $x = (x^0, \vec x)$, $q = (q^0, \vec q)$ and 
${\cal R} J_{\mu}(x) J_{\nu}(0) \equiv  \Theta (x^0) [J_{\mu}(x), J_{\nu}(0)]$
with the conserved vector current $J_\mu$; $\langle \cdots \rangle_{\mu_N, T}$
denotes the grand canonical ensemble average.

We consider vector currents of QCD with isospin quantum numbers of the respective
vector mesons specified as
$J_\mu = \frac 12 ( \bar u \gamma_\mu u \mp \bar d \gamma_\mu d)$, where
the negative (positive) sign is for the $\rho$ ($\omega$) meson and   
$J_\mu = \bar s \gamma_\mu s$ is for the $\phi$ meson current.
Since we focus on the limit $\vec q \to 0$ in the rest frame of the medium,
only the longitudinal invariant 
$\Pi^R_L = {\Pi^{R \, \mu}}_\mu / (- 3 q^2) \vert_{\vec q \to 0}$ of the correlator
(\ref{eq_1}) is needed in our analysis.

Due to the analyticity in the upper half of the complex $q^0$ plane the retarded
correlation function $\Pi^R_L$ satisfies the standard dispersion relation
in a medium
\begin{equation}
\Pi_{L}^{R} (q^0; \mu_N , T) = \frac 1 \pi  
\int_0^\infty d s \; \frac{{\rm Im} \Pi^R_L (s; \mu_N , T)}{s - (q_0 + i \epsilon)^2},
\label{eq_2}
\end{equation}
where  a subtraction has been omitted.
For large $Q^2 \equiv - q_0^2 > 0$ one can evaluate the correlator $\Pi^R_L$
by expanding the product of currents in (\ref{eq_1}) by means of the operator product
expansion (OPE). The result can be written as
\begin{equation}
\Pi^R_L (Q^2) = - C_0 \; \ln Q^2 
+ \sum_{n=1}^{\infty} \frac{C_n}{Q^{2n}},   
\label{eq_3}
\end{equation}
where the quantities $C_n$ include the usual Wilson coefficients and the condensates
as well. It is remarkable that in the considered region 
($q^0 \ne$ real)
the OPE for $\Pi^R_L$ is the same as the OPE for the causal (Feynman) correlator.
This can be used to get an explicit expression of (\ref{eq_3}) in a simple manner.

As usual within the QSR the spectral density
$\rho_{\rm had}(s; \mu_N , T) = \frac 1 \pi {\rm Im} \Pi^R_L (s; \mu_N , T)$ is modeled
by several phenomenological parameters related in particular to the forward scattering
amplitudes of the external current $J_\mu$ and the constituents of the medium.

Performing the Borel transformation with appropriate mass parameter $M^2$
\cite{SVZ} of the dispersion relation eq.~(\ref{eq_2}) and taking into
account the OPE (\ref{eq_3}) one gets the basic equation
for the QSR evaluation in a medium as
\begin{equation}
\int_0^\infty d s \; \rho_{\rm had}(s) \mbox{e}^{-s / M^2} =
M^2 \left( C_0 + \sum_{n=1}^{\infty} \frac{C_n}{(n -1)! \, M^{2n}} \right)
\label{eq_4}
\end{equation}
which formally looks similar to the corresponding equation derived in
\cite{Hatsuda_Koike_Lee} for the Borel sum rules at finite temperature.

\section{OPE and evaluation of the condensates at finite $\mu_N$ and $T$} 

The coefficients $C_n$, which define the r.h.s. of the basic equation 
(\ref{eq_4}), include the Wilson coefficients and the grand canonical ensemble
average of the corresponding products of quark and gluon field operators.
Their general structure up to $n = 3$ is given, for instance, in 
\cite{Hatsuda_Koike_Lee}.
For $\rho$ and $\omega$ mesons one has   
\begin{eqnarray}
C_0
& = & 
\frac{1}{8 \pi^2} \left( 1 + \frac{\alpha_s(\mu^2)}{\pi} \right),
\label{eq_5.0}\\
C_1 
& = &
- \frac{3 m_q^2}{4 \pi^2},
\label{eq_5.1}\\ 
C_2
& = &
m_q \langle \bar u u \rangle_{\mu_N , T} + 
\frac{1}{24} \langle \frac{\alpha_s}{\pi} G^2 \rangle_{\mu_N , T} -
i \frac{4}{3} 
\frac{q^\mu q^\nu}{Q^2}
\langle \hat S \hat T (\bar u \gamma_\mu D_\nu u) \rangle_{\mu_N , T}, 
\label{eq_5.2}\\ 
C_3
& = &
- \frac{\pi \alpha_s}{2}
\langle \left( \bar u \gamma_\mu \gamma_5 \lambda^a u 
\mp \bar d \gamma_\mu \gamma_5 \lambda^a d \right)^2 \rangle_{\mu_N ,
T} 
\nonumber \\
& & -
\frac{\pi \alpha_s}{9}
\langle \left( \bar u \gamma_ \mu \lambda^a u + 
\bar d \gamma_ \mu \lambda^a d \right) 
\sum_{q = u, d, s} \bar q \gamma^\mu \lambda^a q \rangle_{\mu_N , T}
\nonumber \\
& & +
i \frac{16}{3} \frac{q^\mu q^\nu q^\lambda q^\sigma}{Q^4} 
\langle \hat S \hat T \left( \bar u \gamma_\mu D_\nu D_\lambda D_\sigma u  
\right) \rangle_{\mu_N , T}, \label{eq_5.3}
\end{eqnarray}
where again the minus (plus) sign corresponds to $\rho$ ($\omega$) meson;
$\alpha_s (\mu^2)$ is the strong coupling constant which is taken at
the renormalization point $\mu^2 = 1$ GeV$^2$ according to common practice;
$m_q = \frac12 (m_u + m_d)$ stands for the average light quark mass;
$G^2 = G^a_{\mu \nu} G_a^{\mu \nu}$ with the gluon field strength tensor 
$G_a^{\mu \nu}$. The covariant derivative is defined as
$D_\mu = \partial_\mu + \frac i2 g A_\mu^a \lambda^a$ with SU(3) color matrices
$\lambda^a$ normalized as Tr$( \lambda^a \lambda^b) = 2 \delta^{ab}$.
The operator $\hat S \hat T$ creates a symmetric and traceless expression with
respect to the Lorentz indices. 

To get the $\mu_N$ and $T$ dependence of the condensates,
entering the coefficients $C_{2,3}$, we assume that the system,
at small density and temperature, can be described 
by non-interacting nucleons and pions.
For small enough $\mu_N$ and $T$ the particle gas is dilute and the grand canonical
ensemble average of an operator $\hat {\cal O}$ can be approximated by
\begin{eqnarray}
\langle \hat {\cal O} \rangle_{\mu_N , T} 
& = &  
\langle \hat {\cal O} \rangle_0 +
3 \int \frac{d^3 p}{(2 \pi)^3 \, 2 E_p}
n_B  
\langle \pi (p)| \hat {\cal O}| \pi (p) \rangle
\nonumber\\ 
& & + 
4 \int\frac{d^3 k}{(2\pi)^3 \, 2 E_k} 
n_F  
\langle N(k)| \hat {\cal O}| N(k) \rangle,
\label{eq_6}
\end{eqnarray}
where $\langle \hat {\cal O} \rangle_0$ is the vacuum expectation value,
and $n_B = [{\rm e}^{E_p/T} - 1]^{-1}$ 
and $n_F = [{\rm e}^{(E_k - \mu_N)/T} + 1]^{-1}$
are thermal Boson and Fermion distributions; 
we use the covariant normalization of the one-particle pion (nucleon) state:
$\langle \pi (p) \vert \pi (p') \rangle = (2 \pi)^3 \; 2 E_p\;
\delta (\vec p - \vec p\,')$
($\langle N(k) \vert N(k') \rangle = (2 \pi)^3 \; 2 E_k \, 
\delta^3 (\vec k - \vec k')$), where
$E_p = \sqrt{\vec p\,^2 + m_\pi^2}$
($E_k = \sqrt{\vec k^2 + M_N^2}$) with
$m_\pi$ ($m_N$) as pion (nucleon) mass; 
the vacuum is normalized as $\langle 0 | 0 \rangle = 1$. 

Let us consider first the matrix elements of the scalar (Lorentz invariant) operators.
For the vacuum quark and gluon condensates,
$\langle \bar q q \rangle_0$ with $q = u, d, s$ and 
$\langle \frac{\alpha_s}{\pi} G^2 \rangle_0$,
we employ values presented in table~1.
For the vacuum four-quark condensate we adopt the vacuum saturation hypothesis
\cite{SVZ} which is valid in the large-$N_c$ limit ($N_c$ is number of colors)
so that the four-quark condensates become proportional to 
the squares of the quark condensates.
To control a deviation from the vacuum saturation we introduce the parameter
$\kappa_0$ in the following manner
\begin{eqnarray}
\langle (\bar q \gamma_\mu \lambda^a q)^2 \rangle_0
& = &
- \langle (\bar q \gamma_\mu \gamma_5 \lambda^a q)^2 \rangle_0
= 
- \frac{16}{9} \kappa_0 \langle \bar q q \rangle_0^2.
\label{eq_10}
\end{eqnarray}
The case $\kappa_0 = 1$ corresponds obviously to the exact vacuum saturation as
used, for instance, in \cite{Hatsuda_Koike_Lee}.

The pion and nucleon matrix elements of scalar operators which contribute to 
(\ref{eq_6}) do not depend on the particle momenta and, therefore, can be calculated
in the limit of vanishing momenta.
The pion matrix elements of quark operators can be expressed via vacuum condensates
by means of the soft pion theorem \cite{Hatsuda_Koike_Lee}
\begin{equation}
\langle \pi \vert \bar u u \vert \pi \rangle =
\langle \pi \vert \bar d d \vert \pi \rangle =
- \frac{1}{f^2_\pi} \langle \bar u u \rangle_0,
\quad
\langle \pi \vert \bar s s \vert \pi \rangle = 0
\label{eq_11}
\end{equation}
with the vacuum pion decay constant $f_\pi = 93$ MeV.

Among four-quark operators in eq.~(\ref{eq_5.3}), only the term
$(\bar u \gamma_\mu \gamma_5 \lambda^a u -
  \bar d \gamma_\mu \gamma_5 \lambda^a d )^2 \equiv {\cal O}_5^-$,
which contributes to the $\rho$ meson QSR, has a non-vanishing pion
matrix element. Applying the soft pion theorem twice leads to
\begin{equation}
\langle \pi \vert {\cal O}_5^- \vert \pi \rangle = -
\frac{16}{3 f_\pi^2}
\langle ( \bar u \gamma_\mu \gamma_5 \lambda^a u )^2 \rangle_0.
\label{eq_12}
\end{equation} 
Using the vacuum saturation assumption with the corresponding parameter
$\kappa_0$ in eq.~(\ref{eq_10}) one gets
\begin{equation}
\langle \pi \vert {\cal O}_5^- \vert \pi \rangle = -
\frac{256}{27 f_\pi^2} \, \kappa_0 \,
\langle \bar u u \rangle_0^2.
\label{eq_13}
\end{equation} 

The pion expectation value for the scalar gluon condensate is obtained
as usual by employing the QCD trace anomaly of the energy-momentum tensor
\cite{Hatsuda_Koike_Lee,SVZ} and the Gell-Mann - Oakes - Renner relation
\begin{equation}
\langle \pi \vert \frac{\alpha_s}{\pi} G^2 \vert \pi \rangle = -
\frac89 \, m_\pi^2.
\label{eq_14}
\end{equation}

By a comparison of the quark condensate in a nucleon medium obtained
by means of the Feynman - Hellmann theorem, applied to the ground state
of nuclear matter in Fermi gas approximation
\cite{Cohen_Furnstahl_Griegel},
\begin{equation}
\langle \bar q q \rangle_{\mu_N} - \langle \bar q q \rangle_0 = \frac12
\frac{d e}{d m_q},
\quad
e = 4 \int_{ \vert \vec k \vert \le k_F}
\frac{d^3 k}{(2 \pi)^3} \sqrt{\vec k^2 + M_N^2},
\quad k_F = \sqrt{\mu_N^2 - M_N^2} 
\label{eq_15}
\end{equation} 
and eq.~(\ref{eq_6}) taken at $T = 0$
one can get the nucleon matrix element of the scalar quark operator as
\begin{equation}
\langle N \vert \bar q q \vert N \rangle =
\frac{M_N \sigma_N}{m_q},
\quad
q = u, d,
\label{eq_16}
\end{equation}
where $\sigma_N = m_q \, dM_N / d m_q$ is the nucleon sigma term.
Similar steps for the $s$ quark condensate result in
\begin{equation}
\langle N \vert \bar s s \vert N \rangle =
y \, \frac{M_N \sigma_N}{m_q}
\label{eq_17}
\end{equation}
with the parameter $y$ defined via $y \sigma_N = 2 m_s \, dM_N / d m_s$.
Our choice of the numerical values of parameters entering 
eqs.~(\ref{eq_16}, \ref{eq_17}) is given in table~1.

To get a proper ansatz for the nucleon matrix element of the scalar four-quark
condensates we extend the widely used ground state saturation
assumption for the in-medium
four-quark condensate at $T = 0$ in the following way
\begin{equation}
\langle (\bar q \gamma_\mu \gamma_5 \lambda^a q)^2 \rangle_{\mu_N} = -
\langle (\bar q \gamma_\mu \lambda^a q)^2 \rangle_{\mu_N} =
\frac{16}{9} \kappa (n_N) \langle \bar q q \rangle_{\mu_N}^2,
\label{eq_18}
\end{equation}
where the density dependent factor $\kappa (n_N)$ is introduced to control
a deviation from the exact ground state saturation in analogy to the 
vacuum saturation in eq.~(\ref{eq_10}).
As pointed out in \cite{Klingl_Weise_Kaiser} 
the factor $\kappa (n_N)$ reflects the contribution
from the scalar low-energy excitations of the ground state and seems to be
weekly dependent on the nucleon density $n_N$, 
so that one can use $\kappa (n_N) = const$
as first approximation. In this case the density dependence of the 
four-quark condensate appears only via the density dependence of the
quark condensate squared. In linear density approximation it is given by
\begin{equation}
\langle \bar q q \rangle_{\mu_N}^2 = \langle \bar q q \rangle_0^2 +
\langle \bar q q \rangle_0
\langle N \vert \bar q q \vert N \rangle \frac{n_N}{M_N}.
\label{eq_19}
\end{equation}
We go beyond the above first approximation 
and assume here also a possible linear density dependence of $\kappa (n_N)$,
keeping in mind that still $\kappa (n_N) = \kappa_0$ at $n_N = 0$ in accordance
with the vacuum case in eq.~(\ref{eq_10}).
To leading order in the density this assumption can be expressed as correction
to the second term of the r.h.s.\ of eq.~(\ref{eq_19}) by a constant factor
$\kappa_N$. As a result, our parameterization of the four-quark condensate
at $T = 0$ has the form
\begin{equation}
\langle (\bar q \gamma_\mu \gamma_5 \lambda^a q )^2 \rangle_{\mu_N} = 
\frac{16}{9} \langle \bar q q \rangle_0^2 \, \kappa_0 \,
\left( 1 + \frac{\kappa_N}{\kappa_0}
\frac{ \langle N \vert \bar q q \vert N \rangle}{\langle \bar q q \rangle_0 }
\frac{n_N}{M_N} \right) .
\label{eq_20}
\end{equation}
The limit $\kappa_N = \kappa_0 = 2.36$ is
used in \cite{Klingl_Weise_Kaiser}, 
while the parameterization $\kappa_N = 1.4$ and $\kappa_0 = 3.3$ with
$\langle \bar q q \rangle_0 = (- 230 \mbox{MeV})^3$ corresponds
to the choice in \cite{Hatsuda_Lee_Shiomi}. 
Below we vary the parameter $\kappa_N$ to estimate the contributions
of the four-quark condensates to the QSR with respect to the main trends
of the in-medium vector meson mass modification.

The needed ansatz for the nucleon matrix element of the scalar four-quark
condensate can be extracted then from the direct comparison of our
parameterization in eq.~(\ref{eq_20}) and the general expression (\ref{eq_6})
for the condensates via the matrix elements (the latter ones to be taken
at $T = 0$) as
\begin{equation}
\langle N \vert (\bar q \gamma_\mu \gamma_5 \lambda^a q)^2 \vert N \rangle = 
\frac{32}{9} \langle \bar q q \rangle_0 \, 
\langle N \vert \bar q q \vert N \rangle \, \kappa_N.
\label{eq_21}
\end{equation}
Since the matrix element (\ref{eq_21}) does not depend on particle momenta
and temperature it can be employed also with the approximation in
eq.~(\ref{eq_6}) for evaluating the four-quark condensates in the general
case with $T \ne 0$ and $\mu_N \ne 0$.

The gluon condensate in the nucleon is obtained in the same manner as for pions
by using the trace anomaly and eqs.~(\ref{eq_16}, \ref{eq_17})
\begin{equation}
\langle N \vert \frac{\alpha_s}{\pi} G^2 \vert N \rangle = -
\frac{16}{9} M_N M_N^0,
\label{eq_22}
\end{equation}   
where $M_N^0$ is the nucleon mass in the chiral limit; for numerical values
see again table~1.

The pion matrix elements of twist-2 operators, which are
symmetric and traceless with respect to Lorentz indices, can generally be written as  
\cite{Nachtmann} 
\begin{equation}
\langle \pi (p) \vert \hat{\rm S} \hat{\rm T} \bar q \gamma_\mu 
D_{\nu} q \vert \pi (p)\rangle = - i \left(
p_{\mu} p_{\nu} - \frac14 g_{\mu \nu} p^2 \right) A_{2, \pi}^{q} (\mu^2)  
\label{eq_135}
\end{equation} 
for $q = u, d, s$ and 
for mass dimension-4 operators, and 
\begin{eqnarray}
\langle \pi (p) \vert \hat{\rm S} \hat{\rm T} \bar q \gamma_{\mu}   
D_\nu D_\lambda D_\sigma q \vert \pi (p) \rangle 
& = & 
i \Bigg[ p_\mu p_\nu p_\lambda p_\sigma 
+ \frac{p^4}{48} \left( g_{\mu \nu} g_{\lambda \sigma} + g_{\mu \lambda}  
g_{\nu \sigma}
+ g_{\mu \sigma} g_{\nu \lambda} \right) \nonumber \\
& &\hspace{-6.5cm} -\frac{p^2}{8} \left( p_\mu p_\nu g_{\lambda \sigma} + 
p_\mu p_\lambda g_{\nu \sigma} + p_\mu p_\sigma g_{\lambda \nu} 
+ p_ \nu p_\lambda g_{\mu \sigma} + p_\nu p_\sigma g_{\mu \lambda} 
+ p_\lambda p_\sigma g_{\mu \nu} \right)
\Bigg] A_{4,\pi}^q (\mu^2)
\label{eq_140}
\end{eqnarray}
for mass dimension-6 operators, respectively. 
The coefficients $A^q_{2,\pi}$ and $A^q_{4,\pi}$ 
are defined by  
\begin{eqnarray}
A_{i, \pi}^q (\mu^2) 
& = & 2 \int_0^1 d x \; x^{i - 1} 
\left[q_\pi (x, \mu^2) + \bar q_\pi (x, \mu^2) \right] ,
\quad 
i = 2, 4
\label{eq_145}
\end{eqnarray}
where $q_\pi (x,\mu)$ is the quark distribution function inside the pion 
at the scale $\mu^2$. 

The nucleon matrix elements for twist-2 operators of 
mass dimension-4,   
$\langle N(k) \vert \hat{\rm S} \hat{\rm T} \bar q \gamma_\mu D_ 
\nu q \vert N(k) \rangle$, 
and mass dimension-6,  
$\langle N(k) \vert \hat{\rm S} \hat{\rm T} \bar q \gamma_\mu
D_\nu D_\lambda D_\sigma q \vert N (k) \rangle$,   
have the same structure as in eqs.~(\ref{eq_135} - \ref{eq_145}),
but with the quark distribution $q_N (x,\mu)$  inside the nucleon   
at scale $\mu^2$ and
$A_{i, N}^q (\mu^2) = 2 \int_0^1 d x \; x^{i - 1}
\left[q_N (x,\mu^2) + \bar q_N (x,\mu^2) \right].$
Our choice of the quantities $A^q_{i, \pi (N)}$ is listed in table~1.

Using the obtained pion and nucleon matrix elements of the corresponding
operators and performing the grand ensemble average eq.~(\ref{eq_6})
one finally gets the values of the coefficients $C_{2,3}$ entering
the basic equation (\ref{eq_4}) for $\rho$ and $\omega$ mesons as
\begin{eqnarray}
C_2 & = & q_2 + g_2 + a_2, \label{eq_27}\\
C_3 & = & q_4 + a_4, \label{eq_28}\\
q_2 & = & m_q \langle \bar u u \rangle_0 
+ 2 M_N \sigma_N Y I_1^N + \frac34 m_\pi^2 \xi^{\rho, \omega} I_1^\pi 
\label{eq_q2}\\
g_2 & = & \frac{1}{24} \langle \frac{\alpha_s}{\pi} G^2 \rangle_0 
- \frac{4}{27} M_N M_N^0 I_1^N 
- \frac{1}{18} m_\pi^2 I_1^\pi \\  
q_4 & = & - \frac{112}{81} \pi \alpha_s \kappa_0 
\langle \bar u u \rangle_0^2 
\left[ 1 -
\frac{\kappa_N}{\kappa_0} 
\frac{4 M_N \sigma_N}{m_q (-\langle \bar u u \rangle_0)} 
Y I_1^N  
- \frac{36}{7 f_\pi^2} \xi^\rho I_1^\pi \right] \label{eq_31}\\
a_2 & = & M_N^2 \, A_{2,N}^{u+d} \, I_1^N + \frac43 A_{2,N}^{u+d} \, I_2^N 
 + \frac34 m_\pi^2 \, A_{2,\pi}^{u+d}\, I_1^\pi 
 + A_{2,\pi}^{u+d} \, I_2^\pi \label{eq_32}\\
a_4 & = & - \frac53 M_N^4 \,A_{4,N}^{u+d} \, I_1^N 
- \frac{20}{3} M_N^2 \, A_{4,N}^{u+d} \, I_2^N 
- \frac{16}{3} A_{4,N}^{u+d}\, I_3^N \nonumber\\
& & - \frac54 m_\pi^4 \, A_{4,\pi}^{u+d} \, I_1^\pi 
- 5 m_\pi^2 \, A_{4,\pi}^{u+d} \, I_2^\pi - 4 A_{4,\pi}^{u+d} \, I_3^\pi,
\label{eq_a4}
\end{eqnarray}
where $Y = 1$, $\xi^{\rho, \omega} = 1$ for $\rho$ and $\omega$
mesons, $\xi^{\rho} = 1$ for the $\rho$ mesons, 
and elsewhere $\xi^{\cdots} = 0$.   
The sequence of replacements $m_q \to m_s$,
$\langle \bar u u \rangle_0 \to \langle \bar s s \rangle_0$
and $ Y \to y m_s / m_q$,
$A_{i, \pi (N)}^{u + d} \to A_{i, \pi (N)}^s$ holds for the $\phi$ meson. 
The above integrals are
\begin{equation}
I_n^N =  \int \frac{d^3 k}{(2 \pi)^3 E_k} k^{2n-2} \, n_F, 
\quad
I_n^{\pi}  =  \int\frac{d^3 p}{(2 \pi)^3 E_p} p^{2n-2} \, n_B.
\end{equation}

The density and temperature dependence of the scalar quark and gluon
condensates $q_2$ and $g_2$ are exhibited in fig.~1 in \cite{Bormio}.
One observes a striking linear density dependence, while for small temperatures 
there is only a tiny change of the condensate.  
A suitable approximation for  
$\langle \bar q q \rangle_{n_N, T} / \langle \bar q q \rangle_0 $,
$\langle \bar s s  \rangle_{n_N, T} / \langle \bar s s \rangle_0 $ and
$\langle \frac{\alpha_s}{\pi} G^2 \rangle_{n_N, T} / 
\langle \frac{\alpha_s}{\pi} G^2  \rangle_0 $
at small temperatures is $1 - \alpha n/n_0$ with 
$\alpha \approx 0.32$, 0.09 and 0.08,
respectively (we use $n_0 = 0.17$ fm$^{-3}$). 

\section{Hadronic spectral function} 

To get insight for modeling the hadronic spectral density
$\rho_{\rm had} (s; \mu_N, T) = 
\frac1\pi \mbox{Im} \Pi_L^R (s; \mu_N, T)$, which enters the l.h.s.\
of the basic equation (\ref{eq_4}), one can decompose the in-medium correlator
$\Pi_{\mu \nu}^R (q; \mu_N, T)$ in the same manner as in eq.~(\ref{eq_6})
within the dilute gas approximation by restricting on not too large values
of $\mu_N$ and $T$
\begin{eqnarray}
\Pi_{\mu \nu}^R (q; \mu_N, T)
& = &  \Pi_{\mu \nu}^R (q; 0, 0) \nonumber \\
&& + 3 \int \frac{d^3 p}{(2 \pi)^3 \, 2 E_p}
n_B  T^\pi_{\mu \nu} (q; \vec p) 
+ 
4 \int\frac{d^3 k}{(2\pi)^3 \, 2 E_k} 
n_F   T^N_{\mu \nu} (q; \vec k),
\label{eq_36}
\end{eqnarray}
where  $\Pi_{\mu \nu}^R (q; 0)$ is the vacuum expectation and 
\begin{eqnarray}
T^\pi_{\mu \nu} (q; \vec p) & = &
i \int d^4 x \ \mbox{e}^{i q x}
\langle \pi (\vec p) \vert {\cal R} J_\mu(x) J_\nu(0) \vert \pi (\vec p) \rangle,\\
T^N_{ \mu \nu} (q; \vec k) & = &
i \int d^4 x \ \mbox{e}^{i q x}
\langle N (\vec k) \vert {\cal R} J_\mu(x) J_\nu(0) \vert N (\vec k) \rangle
\end{eqnarray}
are the forward scattering amplitudes of the external current $J_\mu$ 
for pion or nucleon, respectively.

For the vacuum QSR, only the first term in eq.~(\ref{eq_36}) 
is obviously non-vanishing.
The corresponding spectral density for a given vector meson channel was successfully
modeled in \cite{SVZ} by means of a resonance peak and a continuum
(perturbative QCD) contribution. In a medium one has to take into account
also the contributions arising from the imaginary parts of the second and third
scattering terms in eq.~(\ref{eq_36}). To model such contributions we consider
only absorption like processes related to the matrix elements 
$\langle \pi \vert J_\mu \vert \pi \rangle$ and   
$\langle N \vert J_\mu \vert N \rangle$.
This is still in accordance with the dilute gas approximation when only the
one-particle states are taken into account. In the limit
$\vec q \to 0$ the above scattering terms, called usually Landau damping terms,
can be calculated exactly (cf.\ \cite{Hatsuda_Lee_Shiomi,BS} for details).
As a result, our parameterization of the hadronic spectral density 
in the vector meson channel $V$ is given by  
\begin{equation}
\rho^V_{\rm had} = \rho^V_{\rm Lan} + \rho^V_{\rm res} + \rho^V_{\rm con}
\label{eq_30}
\end{equation}
with the usual splitting into the Landau term 
$\rho^V_{\rm Lan} = (\rho_{\rm scatt}^{V, \pi} + 
\rho_{\rm scatt}^{V, N}) \delta(s)$ (specified further below),
the resonance part $\rho^V_{\rm res}$ and the continuum contribution
$\rho^V_{\rm con} = C_0 \Theta (s - s_0^V)$. The continuum threshold
$s_0^V$ separates the resonance part
of the spectrum from the continuum part.

Without specifying the resonance part one can define the resonance
mass parameter $m_V$ as the normalized first moment via
\begin{equation}
m_V^2 = \frac{\int_0^{s_0} ds \, s \, \rho^V_{\rm res}(s) e^{-s^2/M^2}}
{\int_0^{s_0} ds \, \rho^V_{\rm res}(s) e^{-s/M^2}}
\label{def_m_V}
\end{equation}
which is evident in the zero-width approximation where
$\rho^V_{\rm res} = F_V \; \delta (s - m_V^2)$. 
One should underline that, according to eqs.~(\ref{eq_4},
\ref{def_m_V}),
only the weighted integrals over $\rho^V_{\rm res}(s)$ enter
the QSR. That means that various shapes, such like the pole
approximation or particular Breit-Wigner parameterizations
or double-peak distributions etc., can deliver the same value
of the mass parameter $m_V$ supposed the corresponding integrals
over $\rho^V_{\rm res}$ are fixed.  

The above functions $\rho_{\rm scatt}^{V, \pi (N)}$ 
entering the Landau damping term in eq.~(\ref{eq_30})
of the spectral density are given for the $\rho$ meson by
\begin{eqnarray}
\rho_{\rm scatt}^{\rho, N}
& = & \frac{1}{48 \pi^2} \int_{4 M_N^2}^{\infty} 
d \omega^2 \, \hat n_F
\sqrt{1 - \frac{4 M_N^2}{\omega^2}} 
\left[ 2 + \frac{4 M_N^2}{\omega^2} \right] ,
\label{eq_40}\\
\rho_{\rm scatt}^{\rho, \pi}
& = & \frac{1}{24 \pi^2} \int_{4 m_{\pi}^2}^{\infty}
d \omega^2 \, \hat n_B
\sqrt{1 - \frac{4 m_{\pi}^2}{\omega^2}} 
\left[ 2 + \frac{4 m_{\pi}^2}{\omega^2} \right]
\end{eqnarray}
and 
$\hat n_F = [{\rm e}^{(\omega - 2 \mu_N)/2 T} + 1]^{-1}$,
$\hat n_B = [{\rm e}^{\omega/2 T} - 1]^{-1}$.
For the $\omega$ meson, $\rho_{\rm scatt}^{\omega, \pi} = 0$ due to symmetry,
and $\rho_{\rm scatt}^{\omega, N} = 9 \rho_{\rm scatt}^{\rho, N}$
due to the different isospin structure of the $\rho$  and $\omega$ mesons
\cite{Abee}.
For the $\phi$ meson the Landau damping for the pion and nucleon gas
is negligible \cite{Asakawa_Ko}.


\section{Evaluation of QCD sum rule} 

Inserting the hadronic spectral function eq.~(\ref{eq_30}) in the
basic equation (\ref{eq_4}), taking the derivative with respect
to $M^{-2}$, and using the definition of the mass parameter $m_V$ in 
eq.~(\ref{def_m_V})
one finally gets the sum rule in the form
\begin{equation}
m_V^2 = M^2 \frac{C_0 \left( 1 - [1 + \frac{s_0^V}{M^2}] {\rm e}^{-s_0^V/M^2} \right) 
- C_2 M^{-4} - C_3 M^{-6}}
{ C_0 \left( 1 - {\rm e}^{-s_0^V/M^2} \right)
+ C_1 M^{-2}     
+ C_2 M^{-4} + \frac12 C_3 M^{-6} 
- (\rho_{\rm scatt}^{V, N} + \rho_{\rm scatt}^{V, \pi} ) M^{-2}},
\label{m_v}
\end{equation}
with the above coefficients $C_{0 \cdots 3}$
from eqs.~(\ref{eq_5.0}, \ref{eq_5.1}, \ref{eq_27} - \ref{eq_a4}).
The identification of $m_V$ with the vector meson mass is suggestive,
however, strictly speaking only valid in the limit of narrow
resonances. Then the pole approximation is applicable and, as
shown in \cite{Hatsuda_Koike_Lee}, the width of the resonance is
calculable from the pole residue $F_V$ afterwards. Having this
in mind we focus on the in-medium change of the mass parameter
$m_V$ which is associated to the vector meson mass.

The quantity $m_V(M^2)$ still needs to be averaged 
within a certain Borel window
$M_{\rm min} \cdots M_{\rm max}$. 
The continuum threshold $s_0^V$ is determined
by requiring maximum flatness of $m_V(M^2)$ within the Borel window.
We use a sliding Borel window determined by the ''10\% + 50\% rule'' 
\cite{eval5,eval10}, i.e.,
the minimum Borel mass $M^2_{\rm min}$ is fixed such that the terms 
of order $O (1/M^6)$ on the OPE side contribute not more than $10 \%$, while  
the maximum Borel mass $M^2_{\rm max}$ 
is determined by the requirement that the continuum part equals to
the resonance contribution in the spectral function.
(A fixed Borel window with
$M_{\rm min} = 0.8$ GeV$^2$ and $M_{\rm max} = 1.5$ GeV$^2$
(2.5 GeV$^2$ for the $\phi$ meson) delivers the same results.)
  
\begin{figure}
\begin{center}
\includegraphics[width=12cm]{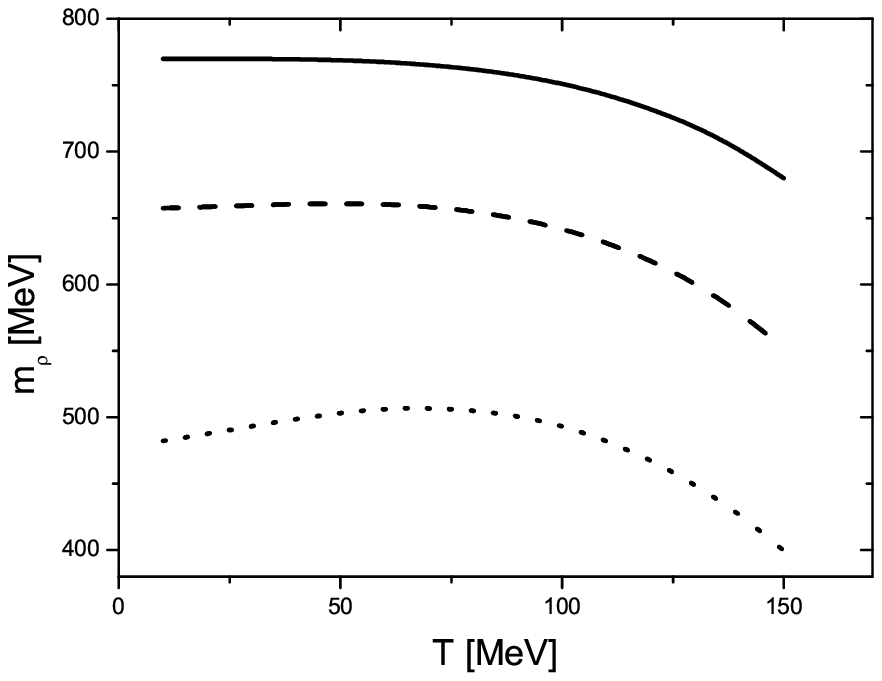}
\includegraphics[width=12cm]{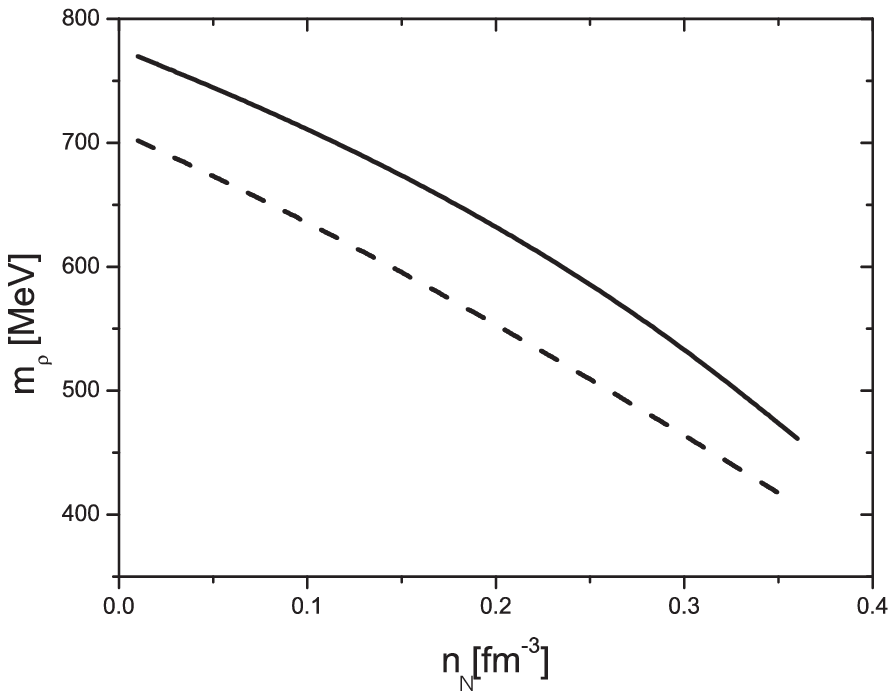}
\caption{The mass parameter $m_\rho$ as a function of temperature
(a, upper panel) and density (b, lower panel).
The solid (dashed, dotted) curves in (a) are for 
densities 0.01 (0.17, 0.34) fm$^{-3}$, while the
solid (dashed) curves in (b) are for
temperatures 20 (140) MeV.}
\label{fig_rho_1}
\end{center}
\end{figure}

\begin{figure}
\begin{center}
\includegraphics[width=12cm]{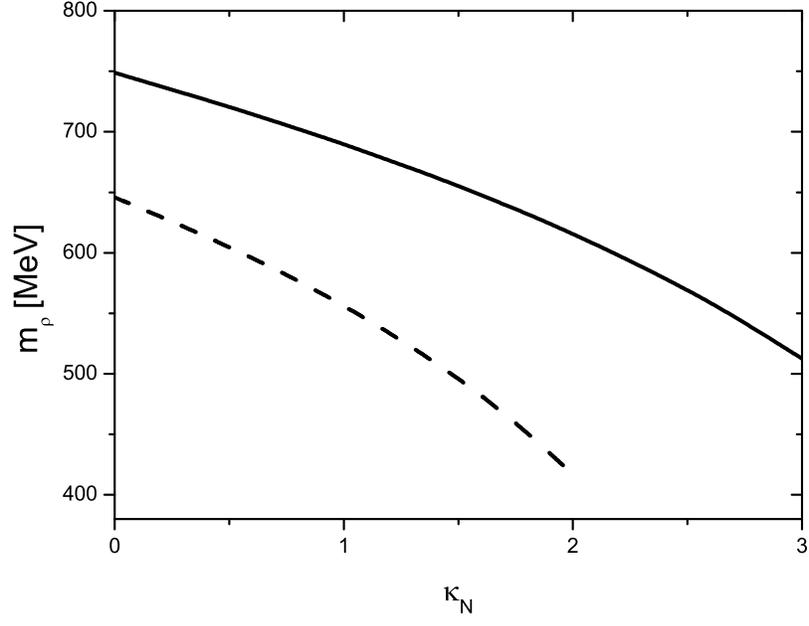}
\caption{The mass parameter $m_\rho$ 
as a function of $\kappa_N$ for $\kappa_0 = 3.0$ and 2.0 
(solid and dashed curves, respectively)
at $n_N = 0.15$ fm$^{-3}$ and $T = 20$ MeV.}
\label{fig_rho_2}
\end{center}
\end{figure}

\begin{figure}
\begin{center}
\includegraphics[width=12cm]{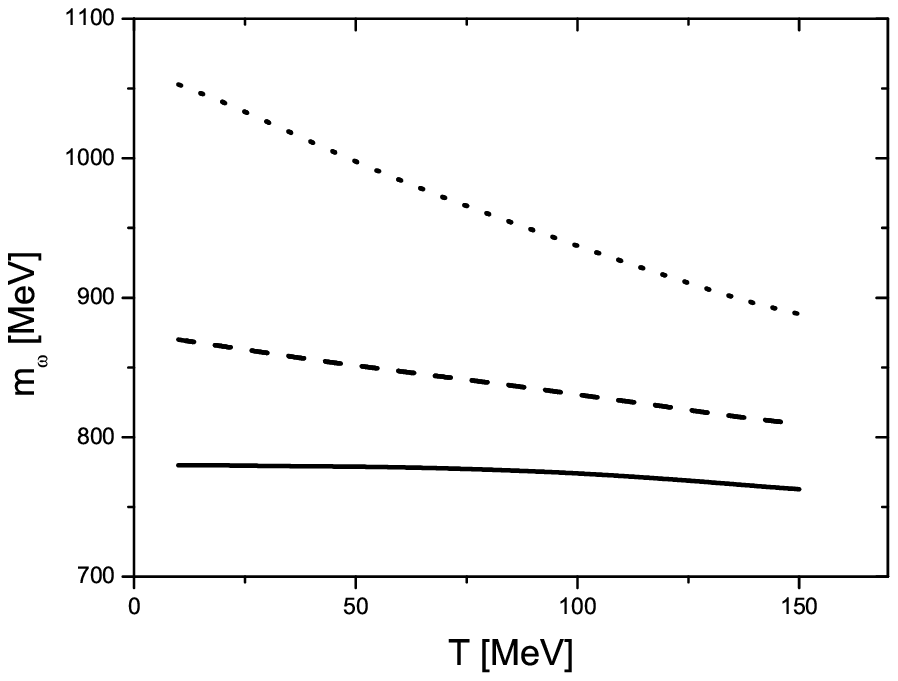}
\includegraphics[width=12cm]{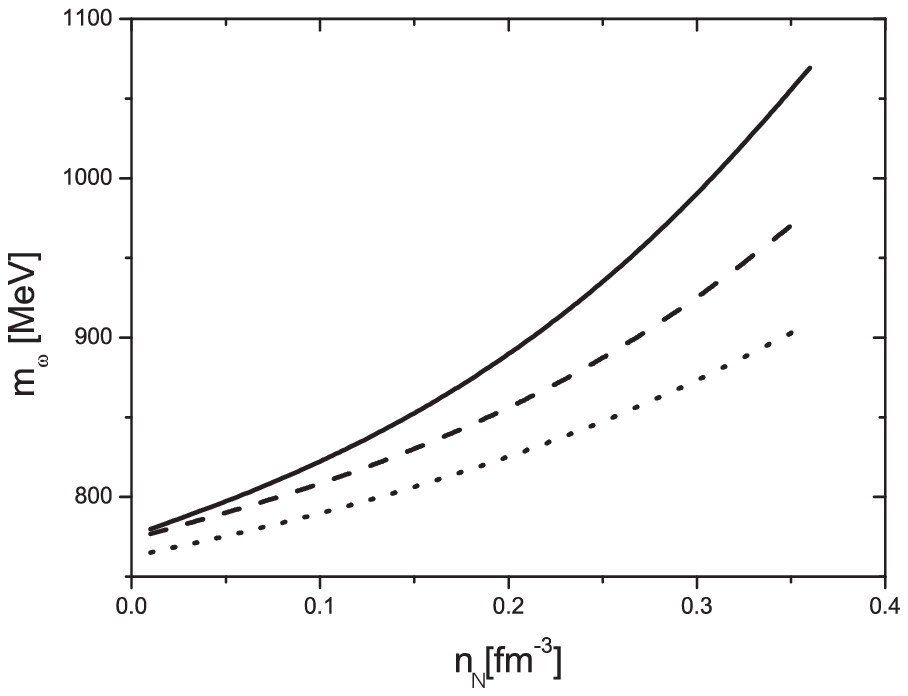}
\caption{As in fig.~\ref{fig_rho_1} but for the $\omega$ meson.
In (b) the temperatures are 20, 80 and 140 MeV (from top to bottom).} 
\label{fig_omega_1}
\end{center}
\end{figure}

\begin{figure}
\begin{center}
\includegraphics[width=12cm]{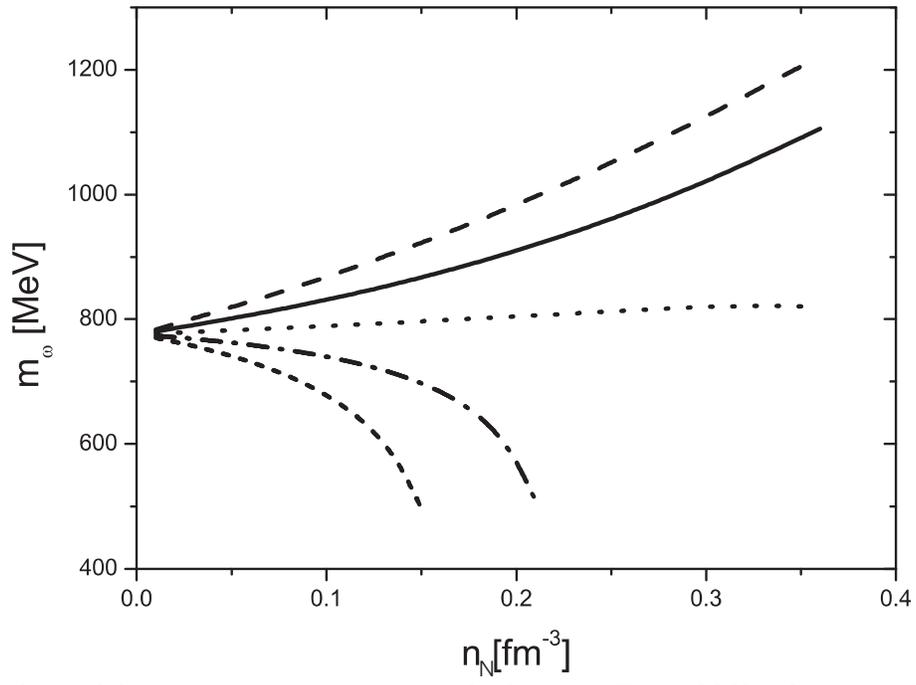}
\caption{The dependence of the $\omega$ meson mass parameter on the
density at $T = 20$ MeV and $\kappa_0 = 3.0$.
The values of $\kappa_N$ are 0, 1, 2, 3, 4 (from top to bottom).} 
\label{fig_omega_2}
\end{center}
\end{figure}

\begin{figure}
\begin{center}
\includegraphics[width=12cm]{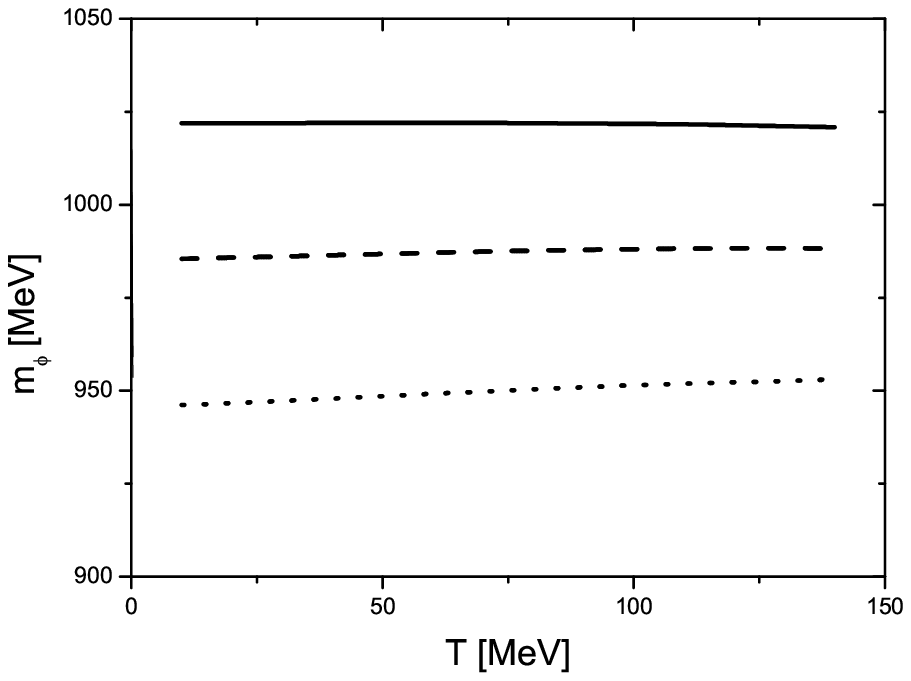}
\includegraphics[width=12cm]{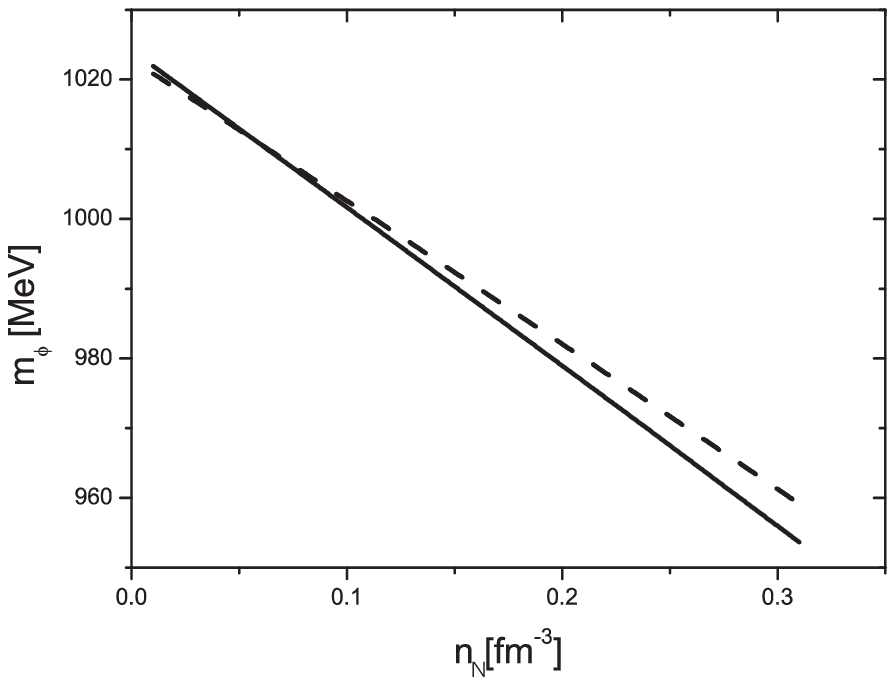}
\caption{As in fig.~\ref{fig_rho_1} but for the $\phi$ meson.} 
\label{fig_phi}
\end{center}
\end{figure}

The results of our QSR evaluation for the density and temperature dependence
of the resonance mass parameter $m_V$ are exhibited in 
figs.~\ref{fig_rho_1} - \ref{fig_phi}.
As seen in figs.~\ref{fig_rho_1}a and b 
the approximately linear dropping of $m_\rho$ with increasing
density appears to be stable with respect to variations of the temperature
in wide region $0 \le T \le 140$ MeV. In particular, if one parameterizes
the four-quark condensate (see eq.~(\ref{eq_20})) by $\kappa_N = 1.4$ and
$\kappa_0 = 3.3$ according to \cite{Hatsuda_Lee_Shiomi}, the density dependence can be
approximated by $m_\rho = m_\rho^0 ( 1 - a_\rho n_N / n_0)$ with 
$a_\rho \approx 0.16$ remaining constant with respect to variations of the
temperature within 30\% accuracy. 
In fig.~\ref{fig_rho_1}a one also observes that at fixed density 
the mass parameter $m_\rho$
suffers only small changes when increasing the temperature up to 100 MeV.
This is in line with the Eletsky-Ioffe mixing theorem on the in-medium
behavior of the vector and axial-vector correlation functions
in the chiral limit \cite{Elitsky_Joffe,Urban,Marco}. 
Up to order $T^2$, at $n_N = 0$
the OPE eq.~(\ref{eq_3}) for the $\rho$ meson correlator in a pion medium
can be rewritten in the following form
\begin{equation}
\Pi_L^R (Q^2; T) = \frac{1}{3 Q^2} \left[
\Pi_\rho (Q^2; 0) + \epsilon \left( \Pi_{a_1} (Q^2; 0) - \Pi_\rho
(Q^2; 0) \right) \right],
\label{eq_34}
\end{equation}
where  
$\Pi_{\rho, a_1} (Q^2; 0) \equiv \Pi^{\mu \mu }_{\rho, a_1} (Q^2; 0)$ 
are the vacuum vector and axial-vector 
correlators and $\epsilon = \frac{T^2}{6 f_\pi^2} B \left( \frac{m_\pi}{T} \right)$
with $B(x) = \frac{6}{\pi^2} \int_x^\infty dy \sqrt{y^2 - x^2} /
(\exp(y) -1)$ (note that $B(0) = 1$). In deriving eq.~(\ref{eq_34})
we use the dilute gas averaging (\ref{eq_6}) and the vacuum saturation
hypothesis (\ref{eq_10}) which obviously give 
$\Pi_{a_1}(0) - \Pi_\rho(0) = \frac{32 \pi \alpha_s}{3 Q^4} 
\langle \bar q q \rangle_0^2 \kappa_0$. From eq.~(\ref{eq_34}) one can
conclude that the $\rho$ meson
spectral function (\ref{eq_30}) used within the framework of the Borel QSR
satisfies at the same time the following equation
\begin{equation}
\int_0^\infty ds \frac{ \rho_{\rm had}^\rho (s; T)}{s + Q^2} 
= \frac{1}{3 Q^2} \left[
\Pi_\rho (Q^2; 0) + \epsilon ( \Pi_{a_1} (Q^2; 0) - \Pi_\rho (Q^2; 0)) \right],
\label{eq_35}
\end{equation} 
so that the in-medium modification of the $\rho$ meson up to order $T^2$ 
stems from the admixture of the vacuum axial-vector correlator, which
is not considerable as long as the temperature is sufficiently small in comparison
with chiral transition temperature.

The general trends of the $\rho$ meson mass dependence on the density in the region
of low temperature, which is relevant for the heavy-ion experiments at HADES,
can be understood by means of the following approximation
relying on the pole approximation of $\rho_{\rm res}^\rho$
\begin{equation}
F_\rho \mbox{e}^{- m_\rho^2 / M^2} \approx
C_0 (1 -  \mbox{e}^{- s_0^\rho / M^2}) M^2
- \rho_{\rm scatt}^{\rho N} +
\frac{a_2}{M^2} +
\frac{q_4}{2 M^4},
\label{eq_45}
\end{equation}
where the four-quark condensate $q_4$,
the non-scalar condensate $a_2$ and 
the Landau damping term $\rho_{\rm scatt}^{\rho N}$
are given by eqs.~(\ref{eq_31}, \ref{eq_32}, \ref{eq_40}).
Since the gluon condensate is weakly dependent on the density, and the 
quark condensate $q_2$ and the non-scalar condensate $a_4$
have rather small contributions, we have omitted the corresponding terms 
in eq.~(\ref{eq_45}). It is clear from eq.~(\ref{eq_45}) that in general
the scattering term tends to increase the mass $m_\rho$, while the
non-scalar condensate $a_2$ tends to decrease the value of $m_\rho$.
When the value of the Borel mass $M^2$ is about in the center of the
Borel window, i.e. $M \sim 1$ GeV, the numerical value of
$\rho_{\rm scatt}^{\rho N}$ is very close to $a_2 / M^2$, so that there
is a cancellation of these terms in a wide range of nucleon densities.
Therefore, the density dependence of $m_\rho$ is determined by the
behavior of the four-quark condensate, in spite of the fact that the
term $q_4 / (2 M^2)$ is much smaller than 
$\rho_{\rm scatt}^{\rho N}$ or $a_2 / M^2$ separately. Actually just the 
linear dependence of the four-quark condensate on the density leads to
the approximate linear decease of $m_\rho$ with increasing density,
as exhibited in fig.~\ref{fig_rho_1}b.

Due to the crucial role of the four-quark condensate for the density dependence
of $m_\rho$ it is useful to vary the parameters $\kappa_0$ and $\kappa_N$
which determine the numerical value of $q_4$ and its dependence on the density
(see eq.~(\ref{eq_31})). As seen in fig.~\ref{fig_rho_2} 
the main trend in decreasing the
mass $m_\rho$, obtained above for the parameterization 
$\kappa_N / \kappa_0 = 0.41$
\cite{Hatsuda_Lee_Shiomi}, is stable with respect to the variation of 
$\kappa_0$ and $\kappa_N$ within the reasonable numerical limits
$2.0 \le \kappa_0 \le 3$ and $0 \le \kappa_N \le 3$.
(Note that the value of $\kappa_0$ is usually adjusted to the vacuum mass of the
$\rho$ meson and the ratio $\kappa_N / \kappa_0$ is restricted by the
conditions $q_4 < 0$ and $q_4 \to 0$ with increasing density and temperature.)
In accordance with eq.~(\ref{eq_31}) for $q_4$ and eq.~(\ref{eq_45})
the increase of the ratio
$\kappa_N / \kappa_0$ leads to a stronger shift of $m_\rho$ at fixed density,
as displayed in fig.~\ref{fig_rho_2}. 
This further confirms the crucial role of the four-quark
condensate for the in-medium behavior of the $\rho$ meson mass.

In a sufficiently dense nucleon medium one can expect
a strong broadening of the $\rho$ meson width due to inelastic $\rho N$
scattering \cite{Rapp_Wambach,Klingl_Weise_Kaiser}.
In principle, the inelastic processes should be included in the 
hadronic spectral function (\ref{eq_30}), but in this case the
OPE side should also include the higher order effects related,
in particular, to two-pion states. Since such effects are beyond
the often used dilute gas approximation we do not concentrate
here on the in-medium modifications of the vector meson widths and
insist to postpone this to a separate and self-consistent study with
higher order effects in the OPE side for finite density and temperature. 

The results of our QSR evaluation of the density and temperature dependence
of the $\omega$ meson mass are exhibited in figs.~\ref{fig_omega_1}a
and b. 
Due to the comparatively large scattering term 
$\rho_{\rm scatt}^{\omega N} \gg \rho_{\rm scatt}^{\rho N}$
the density and temperature behavior of the $\omega$ meson mass differs
essentially from the corresponding in-medium modifications of the 
$\rho$ meson. In particular, for the parameterization of the four-quark
condensate with $\kappa_N / \kappa_0 = 0.41$, which is used above for the
$\rho$ meson, the mass of the $\omega$ meson increases with increasing
density, as shown in fig.~\ref{fig_omega_1}b for a wide region of the temperature.    
This can be still understood within the approximation (\ref{eq_45}), but
now one has to take into account that the numerically large term
$\rho_{\rm scatt}^{\omega N}$ overwhelms the contributions from the non-scalar
condensate $a_2$.
A similar behavior of the $\omega$ meson mass was obtained in 
\cite{Abee} for the case $T = 0$.

We consider also the sensitivity of the density
dependence of the $\omega$ meson with respect to various parameterizations
of the $q_4$ term. In fig.~\ref{fig_omega_2} 
we plot the density dependence of the $\omega$
meson mass at fixed temperature $T = 20$ MeV for various values of the parameter
$\kappa_N$ which reflects the strength of the density dependence of the four-quark
condensate. As seen in fig.~\ref{fig_omega_2}
the in-medium modification of the $\omega$ meson
changes even qualitatively under variation of the parameter $\kappa_N$
(at fixed $\kappa_0 = 3.0$):
for $\kappa_N \stackrel{<}{\sim} 2$ 
the $\omega$ meson mass increases with
increasing density, while for $\kappa_N > 2$ it drops.
Such a drastic change of the density dependence happens since 
the behavior of the density
dependent part of the combination 
$- \rho_{\rm scatt}^{\omega N} + a_2 / M^2 + q_4 / (2 M^4)$
in the QSR eq.~(\ref{eq_45}) is governed by the parameter $\kappa_N$.
The obtained strong sensitivity of $m_\omega$ on the four-quark
condensate holds also for higher temperature. From the above considerations
one can conclude that the sign of the in-medium $\omega$ mass shift is
directly related to the density dependence of the four-quark condensate.

In contrast to the $\rho$ and $\omega$ mesons the density dependence of the
$\phi$ meson is governed mainly by the in-medium chiral condensate of strange
quarks $m_s \langle \bar s s \rangle_{\mu_N, T}$, which has the dominant
contribution in the OPE. As exhibited in figs.~\ref{fig_phi}a and b, 
$m_\phi$ drops with increasing nucleon density. 
(To adjust the vacuum value of the $\phi$ meson mass we employ
$\kappa_0 = 0.6$ and the values of parameters as listed in table~1.) 
This is in agreement with results of
\cite{Hatsuda_Lee} obtained at $T = 0$. Due to the weak dependence of the
chiral quark condensate on temperature the linear decrease
of $m_\phi$ with increasing nucleon density holds in a wide range of temperatures
up to 140 MeV. We also find a very tiny change of $m_\phi$ 
by variations of the four-quark condensate due to the density dependence.
From the above analyses one can conclude that just the measurement
of the $\phi$ meson mass shift in a nucleon medium is most appropriate
to search for in-medium modifications of the chiral quark condensate.
In addition, the $\phi$ meson is expected to keep its quasi-particle
character as narrow resonance \cite{Klingl_Weise_Kaiser}. Such measurements
can be accomplished at HADES in heavy-ion experiments along with
pion or proton induced $\phi$ production at nuclei. 

\section{Summary} 

In summary we present for the first time a systematic analysis of the 
QCD Borel sum rules for light vector mesons at finite nucleon density 
\underline{and} temperature. Our approach is based on the dilute gas
approximation for a nucleon medium which is appropriate at not too
large nucleon density (say, $n < 2 n_0$) and temperature
(say, $T < 100$ MeV). Such conditions are expected to be reached
in heavy-ion experiments at HADES. We find that in-medium modifications
of the $\rho$ and $\omega$ mesons masses are dominated, within the Borel QSR
approach, by the dependence of the four-quark condensate 
$\langle (\bar q q )^2 \rangle_{\mu_n, T}$ on density and temperature.
In particular, the numerical value of the parameter $\kappa_N$,
which describes the strength of the linear density dependence of 
$\langle (\bar q q )^2 \rangle_{\mu_n, T}$ governs the decrease of the
$\rho$ meson as a function on the density. 
At the same time an experimental identification of the $\rho$ meson mass
shift is closely related to the problem of the strong broadening
of its in-medium width. This needs separate considerations
with the Borel QSR including higher density and temperature effects
into the OPE side.
(For a possible strategy to get access to a strongly broadened $\rho$ meson
by means of the double differential $e^+ e^-$ spectra
we refer the interested reader to \cite{KP}.)

For the $\omega$ meson the sign of the in-medium mass shift is changed
by variation of the parameter $\kappa_N$. This also points to the
crucial role of the poorly known four-quark condensate and makes definite
predictions difficult. A direct measurement of the in-medium mass
shift can give, within the considered framework, an important
information on the in-medium behavior of the four-quark condensate.
Since the difference of the vector and axial-vector correlation
functions is proportional to the four-quark condensate the measured
sign of the vector meson mass shift can serve, via the four-quark
condensate, as a tool for determining how fast the nucleon
system approaches the chiral symmetry restoration with increasing
density.

The in-medium 
$\phi$ meson mass shift is directly related to the behavior of the chiral
strange quark condensate $m_s \langle \bar s s \rangle_{\mu_N, T}$
which decreases with density almost independently of the temperature
in a wide region of density and temperature. This offers the chance
to measure the density dependence of the chiral quark condensate via
an in-medium mass shift of the $\phi$ meson in heavy-ion experiments,
supposed the strangeness factor $y$ is not too small
\cite{y}.
Since at nuclear saturation density the mass shift is already noticeable,
one can also complementary search for a $\phi$ meson mass shift 
in hadron - nucleus reactions at the HADES detector
under suitable kinematical conditions. 
  
\subsection*{Acknowledgments}

We tank H.W. Barz, E.G. Drukarev, L.P. Kaptari, R. Hofmann
and G. Zinovjev for useful discussions
O.P.P. acknowledges the warm hospitality of the nuclear theory group
in the research center Rossendorf.
This work is supported by BMBF 06DR921, 
STCU 15a,
CERN-INTAS 2000-349,
NATO-2000-PST CLG 977 482.
   


\newpage
\begin{table}[h!]
\begin{center}
\begin{tabular}{|l|l|c|}
\hline
parameter & numerical value & reference\\[1pt]
\hline
$\alpha_s$ & $0.38$ & \cite{pdb}\\[1pt]
\hline
$m_q$ & $0.0055$ GeV & \cite{Cohen_Furnstahl_Griegel}\\[1pt]
\hline
$m_s$ & $0.130$ GeV & \cite{Cohen_Furnstahl_Griegel}\\[1pt]
\hline
$f_{\pi}$ & $0.093$ GeV &\cite{Cohen_Furnstahl_Griegel}\\[1pt]
\hline
$M_N$ & $0.938$ GeV & \cite{pdb}\\[1pt]
\hline
$M_N^0$ & $0.770$ GeV & \cite{Hatsuda_Lee} \\[1pt]
\hline
$\sigma_N$ & $0.045$ GeV &\cite{Cohen_Furnstahl_Griegel}\\[1pt]
\hline
$y$ & 0.22 & \cite{Hatsuda_Lee,Cohen_Furnstahl_Griegel}\\[1pt]
\hline
$m_{\pi}$ & $0.138$ GeV &\cite{Cohen_Furnstahl_Griegel} \\[1pt]
\hline
$\langle \bar u u \rangle_0 = \langle \bar d d \rangle_0$ & $(-0.245
\, {\rm GeV})^3$ & cf. \cite{Hatsuda_Lee} \\[1pt]
\hline
$\langle \bar s s \rangle_0 $ &  $ 0.7 \langle \bar u u \rangle_0$ & 
\cite{Cohen_Furnstahl_Griegel_Jin} \\[1pt]
\hline
$\langle \frac{\alpha_2}{\pi} G^2 \rangle_0 $ & $(0.33 \, {\rm GeV})^4 $& 
\cite{Cohen_Furnstahl_Griegel_Jin} \\[1pt]
\hline
$A_{2,N}^{u+d} $ & $1.02$ & \cite{Hatsuda_Lee}\\[1pt]
\hline
$A_{4,N}^{u+d} $ & $0.12$ & \cite{Hatsuda_Lee}\\[1pt]
\hline
$A_{2,\pi}^{u+d} $ & $0.97$ &\cite{Hatsuda_Koike_Lee} \\[1pt]
\hline
$A_{4,\pi}^{u+d} $ & $0.255$ &\cite{Hatsuda_Koike_Lee} \\[1pt]
\hline
$A_{2,N}^{s} $ & $0.1$ &\cite{Hatsuda_Lee} \\[1pt]
\hline
$A_{4,N}^{s} $ & $0.004$ & \cite{Hatsuda_Lee} \\[1pt]
\hline
$A_{2,\pi}^{s} $ & $0.08$ &\cite{Asakawa_Ko} \\[1pt]
\hline
$A_{4,\pi}^{s} $ & $0.008$ &\cite{Asakawa_Ko}\\[1pt]
\hline
\end{tabular}
\end{center}
\caption{Set of parameters used. }
\label{tab1}
\end{table}


\newpage

\newpage
\begin{thebibliography}{99}
\bibitem{Weinberg} 
S. Weinberg, Phys. Rev. Lett. {\bf 18}, 507 (1967);\\
J. Kapusta, E.V. Shuryak, Phys. Rev. D {\bf 49}, 4694 (1994).
\bibitem{Elitsky_Joffe} 
M. Day, V.L. Elitsky, B.L. Joffe, Phys. Lett. B {\bf 252}, 620 (1990).
\bibitem{Karsch}  
F. Karsch, E. Laermann, A. Peikert, Nucl. Phys. B {\bf 605}, 579 (2001);\\
F. Karsch, Nucl. Phys. A {\bf 698}, 199 (2002). 
\bibitem{Brown_Rho} G. E. Brown, M. Rho, 
Phys. Rev. Lett. {\bf 66}, 2720 (1991);\\
C. Adami, G.E. Brown, Phys. Rep. {\bf 234}, 1 (1993).
\bibitem{Hatsuda_Lee} 
T. Hatsuda, S.H. Lee, Phys. Rev. C {\bf 46}, R34 (1992).
\bibitem{Hatsuda_Lee_Shiomi} 
T. Hatsuda, S.H. Lee, H. Shiomi, Phys. Rev. C {\bf 52}, 3364 (1995).
\bibitem{HADES} 
http://www-hades.gsi.de.
\bibitem{CERES} 
G. Agakichiev et al. (CERES collaboration), 
Phys. Rev. Lett. {\bf 75}, 1272 (1995);\\
J.P. Wurm et al. (CERES collaboration), 
Nucl. Phys. A {\bf 590}, 103c (1995).
\bibitem{Rapp_Wambach} 
R. Rapp, J. Wambach, Adv. Nucl. Phys. {\bf 25}, 1 (2000).
\bibitem{Gallmeister} 
K. Gallmeister, B. K\"ampfer, O.P. Pavlenko, 
Phys. Lett. B {\bf 473}, 20 (2000).
\bibitem{BS} 
A. I. Bochkarev, M. E. Shaposhnikov, 
Phys. Lett. B {\bf 145}, 276 (1984);
Nucl. Phys. B {\bf 268}, 220 (1986).
\bibitem{Hatsuda_Koike_Lee} 
T. Hatsuda, Y. Koike, S.H. Lee, 
Nucl. Phys. B {\bf 394}, 221 (1993).
\bibitem{SVZ} 
M.A. Shifman, A.I. Vainshtein, V.I. Zakharov, 
Nucl. Phys. B {\bf 147}, 385 (1979); ibid p.\ 448; ibid p.\ 519.
\bibitem{Cohen_Furnstahl_Griegel} 
T.D. Cohen, R.J. Furnstahl, D.K. Griegel, 
Phys. Rev. C {\bf 45}, 1881 (1992).
\bibitem{Klingl_Weise_Kaiser} 
F. Klingl, N. Kaiser, W. Weise, 
Nucl. Phys. A {\bf 624}, 527 (1997).
\bibitem{Nachtmann} 
O. Nachtmann, Nucl. Phys. B {\bf 63}, 237 (1973);\\
H. Georgi, H.D. Politzer, Phys. Rev. D {\bf 14}, 1829 (1976).
\bibitem{Bormio} S. Zschocke, B. K\"ampfer, O.P. Pavlenko,
nucl-th/0202066. 
\bibitem{Abee} 
A.K. Dutt-Mazumber, R. Hofmann, M. Pospelov, 
Phys. Rev. C {\bf 63}, 015204 (2000).
\bibitem{Asakawa_Ko} 
M. Asakawa, C.M. Ko, Nucl. Phys. A {\bf 572}, 732 (1994).
\bibitem{eval5} 
S. Leupold, W. Peters, U. Mosel, 
Nucl. Phys. A {\bf 628}, 311 (1998).
\bibitem{eval10} 
X. Jin, D.B. Leinweber, Phys. Rev. C {\bf 52}, 3344 (1995).
\bibitem{Urban} 
M. Urban, M. Buballa, J. Wambach, 
Phys. Rev. Lett. {\bf 88}, 042002 (2002).
\bibitem{Marco} 
E. Marco, R. Hofmann, W. Weise, Phys. Lett. B {\bf 530}, 88 (2002).
\bibitem{KP} 
B. K\"ampfer, O.P. Pavlenko, Eur. Phys. J. A {\bf 10}, 101 (2001).
\bibitem{y} 
M. Baylac (for the HAPPEX collaboration) 
Nucl. Phys. A {\bf 680}, 269 (2000);\\
K.F. Liu, J. Phys. G {\bf 27}, 511 (2001);\\
C. Micheal, C. McNeile, D. Hepburn, 
Nucl. Phys. Proc. Suppl. {\bf 106}, 293 (2002). 
\bibitem{pdb} 
D.E. Groom et al. (Particle Data Group) 
Europ. Phys. Journ. C {\bf 15}, 1 (2000).
\bibitem{Cohen_Furnstahl_Griegel_Jin} 
T.D. Cohen, R.J. Furnstahl, D.K. Griegel, X. Jin, 
Prog. Part. Nucl. Phys. {\bf 35}, 221 (1995). 
\end{thebibliography}
\end{document}